# Controlling the emission from semiconductor quantum dots using ultra-small tunable optical microcavities.

Ziyun Di, Helene V. Jones, Philip R. Dolan, Simon M. Fairclough, Matthew B Wincott, Johnny Fill, Gareth M. Hughes, and Jason M. Smith*

Department of Materials, University of Oxford, Parks Road, Oxford OX1 3PH, UK

**Abstract**

We report the control of spontaneous emission from CdSe/ZnS core-shell quantum dots coupled to novel open-access optical microcavities. The cavities are fabricated by focused ion beam milling, and provide mode volumes less than a cubic micrometre. The quantum dot emission spectrum, spatial modes, and lifetime are all modified substantially by the presence of the cavity, and can be tuned by actively varying the cavity length. An increase in emission rate of 75% is achieved at room temperature, attributed to the Purcell effect in the 'bad emitter' regime. We demonstrate a high degree of control over the emission from the dots, including near single-mode operation and the ability to detect strong emission from individual nanocrystals.

*Corresponding author: jason.smith@materials.ox.ac.uk

**Introduction**

Controlled coupling between electric dipole transitions in matter and the surrounding electromagnetic field is a central theme of modern optoelectronics. The coupling of nanomaterials such as semiconductor quantum dots to discrete modes in optical microcavities and nanocavities has been researched extensively in recent years, opening the door to such devices as displays, sensors,[1] nanolasers,[2] and single-photon sources for metrology and quantum information technologies. [3,4,5] Most of these applications operate in the so-called 'weak coupling regime' of cavity quantum electrodynamics, whereby leakage of light from the cavity occurs faster than the interaction rate between the emitting dipole and cavity mode.[6] In this regime, spontaneous emission can be channeled into the desired cavity modes by modification of the local density of states of the electromagnetic field (the Purcell effect[7]). The decay rate, emission spectrum and spatial distribution of emitted photons can each be tailored by the appropriate choice of cavity.

For other applications such as quantum computing, the strong coupling regime is desirable, where the cavity leakage rate is slower than the coupling strength and natural decay rate of the atom. In this regime, coherent exchange of information between electronic and optical states is possible, and highly non-linear effects permit the construction of quantum logic gates. [8,9,10]

One design of microcavity that has been the subject of significant effort in recent years is the open-access microcavity, a Fabry-Pérot resonator with opposing mirrors on separate substrates, and with one or both mirrors concave in shape to provide lateral confinement of the cavity mode.[11,12,13,14,15] The attractions of this design are that the intensity maximum of the mode is accessible for coupling to free-standing objects such as atoms, molecules, and nanoparticles; they are fully wavelength tunable *in situ* by controlling the cavity length with a piezoelectric actuator; and the leakage of light from the cavity mode through the mirrors can be coupled efficiently into a Gaussian beam or optical fibre waveguide.[15] A few different approaches have been used to fabricate such cavities,[11,12,13,14] but recently ablation using a carbon dioxide laser has been the favoured method, producing radius of curvature as low as 40 μm and with surface roughness as low as 1Å.[15] Purcell enhancement of the emission rate from quantum dots at low temperature has also been achieved.[16]

Here we report on our latest experiments with open access microcavities fabricated by ion beam milling.[17] By producing high quality concave surfaces with radius of curvature, $β$, as short as 7 μm combined with cavity lengths, $L$, of 1.6 μm we are able to create the smallest mode volumes for this design of cavity to date, down to 0.53 μm$^3$, whilst retaining quality factors of several thousand. To illustrate the benefits of these small sizes we demonstrate Purcell enhancement of the emission from semiconductor nanocrystal quantum dots (NQDs) at room temperature. This corresponds to the 'bad emitter' regime of cavity quantum electrodynamics, in which dephasing and/or spectral drift of the electronic transition render the transition line width much greater than the cavity line width, with the result that the dot-cavity coupling strength is greatly reduced. To observe Purcell enhancement in this regime the use of ultra-small cavities therefore becomes essential.[18]

Recent theoretical work has shown that the bad emitter regime may be useful for producing advanced room temperature single photon sources[19] and single emitter lasers,[20] by making use of the cavity feeding effect to produce controlled radiation from strongly dephasing solid state photon emitters. A few other recent studies have investigated the room temperature Purcell regime using photonic crystal, Bragg pillar, and whispering gallery mode resonators.[21,22,23,24]

**Experimental**

For our intracavity emitter we use commercially available CdSe/ZnS NQDs (eBioscience). These provide high fluorescence quantum yield ($\eta > 0.9$ [25]), well characterised lifetimes, and are solution-based for ease of introduction into the cavities. The cavities are arrays of fully tunable half-symmetric open-access microcavities similar to those reported previously[17]. For the experiments reported here we used cavities with radius of curvature $\beta = 7$ μm and $\beta = 25$ μm. The mirrors consist of ten pairs of $SiO_2$ ($n = 1.45$) / $ZrO_2$ ($n = 2.095$) with reflectivity $R = 99.4\%$, terminated with a $\lambda/2$ $SiO_2$ layer. PL experiments were carried out using the apparatus depicted in figure 1a. The lower featured mirror of the cavity pair was supported on a piezoelectric actuator to control the cavity length, and microscopy was performed through the fixed upper, planar mirror. A solution of the quantum dots in octadecene ($n = 1.44$) was introduced into the cavity during assembly.

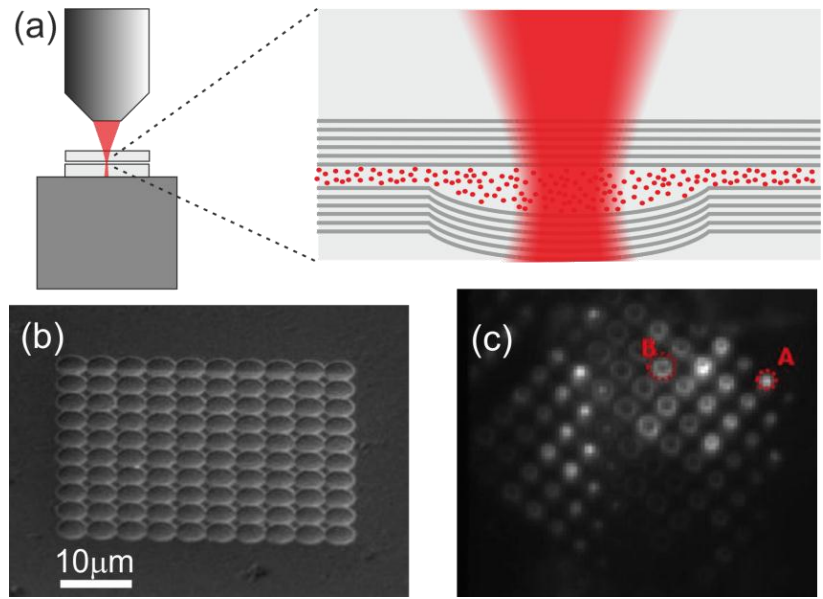

Fig. 1 (a) Schematic of the experimental setup for photoluminescence of NQDs in the half-symmetric microcavities. The upper mirror is fixed and the lower mirror is mounted on a piezoelectric actuator to control the cavity length, (b) electron micrograph of an array of concave features milled into silicon, (c) PL image of the cavity array with a gradient in the cavity length. The two cavities marked A and B provide the emission spectra labeled in figures 2 c and d.

Optical emission from the NQDs in the cavities was investigated using a standard photoluminescence (PL) apparatus. Emission from the NQD ensemble was excited using a laser emitting at 473 nm (PicoQuant LDH470), focused to the cavity using a low magnification objective lens (20X, N.A. = 0.45). The PL signal was collected by the same objective lens and guided to a spectrometer (Acton SP500i) with a liquid nitrogen-cooled CCD (Princeton Spec10). For experiments with very low concentrations of nanocrystals a white light source was installed between the piezo actuator and the lower mirror so that cavity modes could be observed by monitoring the transmission spectrum of the cavities. All measurements were performed at room temperature.

**Results and discussion**

Figure 1(c) shows an image of the PL intensity recorded from a 10 × 10 array of cavities under defocused laser excitation. Much stronger emission can be seen from the individual cavities than from the spaces between them or between the arrays, indicating that the fully confined modes of the half-symmetric cavities are more effective at directing light into the collection optics than are the 1-D confined modes of planar-planar cavities. We will return to the subject of cavity coupling efficiency in more detail later in the paper. An interesting feature of Figure 1(c) is the periodic variation in the spatial mode of the emission from left to right across the image. This is a result of a gradient in the cavity length due to a misalignment of the mirrors by about 5 mrad. We find that within the mode stability criterion of $L < \beta$, the behaviour of individual cavities is extremely robust to such misalignment, with no noticeable effect on the quality of the observed modes.

Figure 2 shows PL spectra taken from four representative cavities with different $\beta$ and L. For comparison, the emission spectrum of the NQD ensemble without the cavity is also shown (Fig. 2(a)). Fig. 2(b) shows a spectrum from a $\beta = 25$ μm cavity with $L = 5.5$ μm, revealing the Hermite-Gauss mode structure of the cavities. Longitudinal modes (TEM$_{00}$) are seen at 614 nm, 640 nm and 667 nm, while TEM$_{mn}$ modes with m + n = 1, 2, 3, 4, and 5 are visible at the short wavelength side of the longitudinal modes at 640 nm and 667 nm respectively. Reducing $L$ to about 1.6 μm increases the free spectral range to about 95 nm in Fig.2(c) whereby only one longitudinal mode is visible within the stop band of the dielectric mirrors (Note that here we define $L$ to include the penetration depth of the field into the mirrors). This spectrum corresponds to cavity A labeled in figure 1(c). Cavity B corresponds to a detuning of about half the free spectral range and reveals the spectrum in

Fig 2(d). The "doughnut" shape PL image observed comes about because each photon couples to multiple modes which interfere destructively for all but the largest radius. Fig. 2(e) shows a near single mode spectrum obtained with a $\beta = 7$ μm cavity with $L = 1.6$ μm.

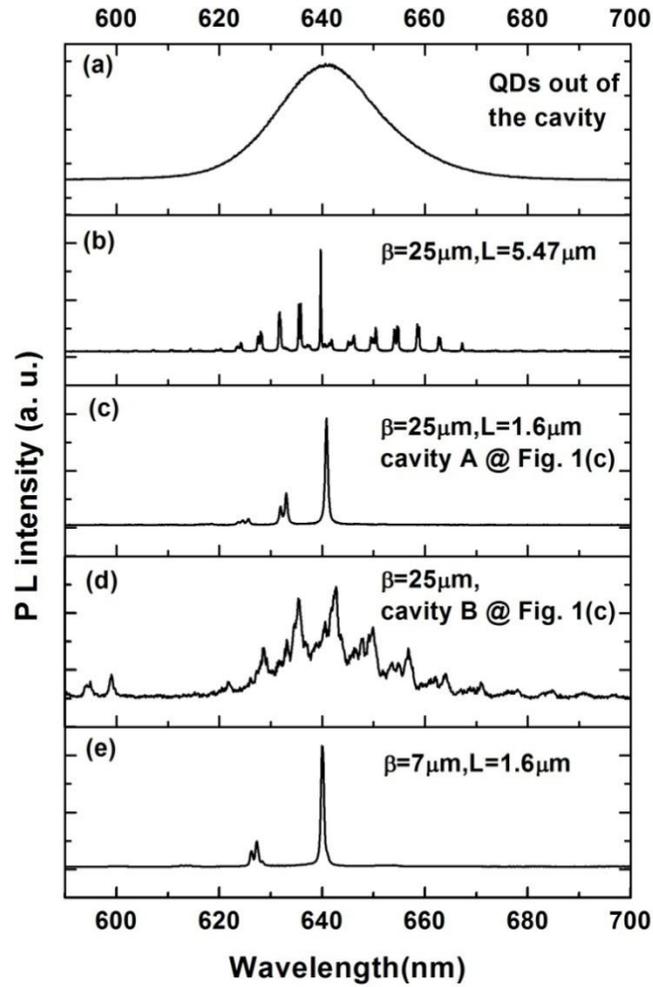

Fig. 2 PL spectra from (a): NQDs out of the cavity as a reference, (c) and (d): NQDs in cavity A and B from Fig. 1(c) respectively; (b) and (e): NQDs in cavity of $\beta = 25$, 7 μm and L = 5.5, 1.6μm respectively.

The luminescence decays of the combined NQD-cavity systems were measured using time-correlated single photon counting with a 100 ps duration excitation pulse and a 900 ps overall timing resolution. Care was taken to keep the excitation intensity below the level where biexcitons can be generated resulting in Auger-limited decay at short time scales.

Figure 3a shows the fluorescence decay for nanocrystals in the smallest cavities, compared to those outside the cavities, but in a similar average dielectric environment. The latter revealed approximately single exponential decays of lifetime 14 ns. The intra-cavity decay, by contrast, is highly non-exponential as a result of the spatial distribution of nanocrystals within the cavity, resulting in a range of coupling strengths. In order to identify the decay rate for nanocrystals with near optimal coupling to the modes (*i.e.* situated at the antinode of the electric field and with the transition dipole aligned parallel to the electric field) we fitted single exponentials only to the first 20 ns of the decay.

Figure 3(b) shows the fitted cavity coupled decay rates relative to the free space decay rate, $\gamma'/\gamma_0$, plotted as a function of mode volume. Data from cavities of two different radii of curvature are included in the figure, each of which provides a series of data points as the cavity is shortened in half-wavelength steps to bring successive $TEM_{00}$ modes to the 640 nm centre wavelength of the quantum dot emission. The cavity length $L$ is established in each case from the measured free spectral range.

Effective mode volumes are calculated using both the analytic expression in the paraxial approximation,

$$V = \frac{\lambda L^2}{4n}\sqrt{\left(\frac{\beta}{L}-1\right)} \qquad [1]$$

and numerically by integrating over the field energy distribution

$$V = \frac{\int_{vol} \varepsilon_r(\underline{r})|E(\underline{r})|^2 d^3\underline{r}}{\varepsilon_r(\underline{r}_{QD})|E(\underline{r}_{QD})|^2} \quad [2]$$

where the electric field distribution $E(\underline{r})$ is established using finite difference time domain (FDTD) modelling software (Lumerical), and $\underline{r}_{QD}$ is the location of the emitting quantum dot. These calculation methods are found to agree well provided that $L < 0.7\beta$ (to satisfy the paraxial approximation), using a field penetration depth into each mirror of 1.08 λ,[26] and so the analytic method is used for all subsequent analysis unless otherwise stated.

The experimental decay data reveal that a change in the recombination rate can only be observed for mode volumes $V < 2$ μm³ whereupon the rate increases steeply to 1.75 $\gamma_0$ at $V = 0.53$ μm³. This is the first time to our knowledge that the dependence of decay rate on mode volume has been mapped directly.

The data shown are from cavities with two different radii of curvature, yet the graph suggests that only the mode volume is important in determining the exciton lifetime. This mode volume dependence is a distinctive characteristic of the bad emitter regime for these cavities, in which the effective Q factor is determined by the emitter and not the cavity. In the well-known analytic expression for the Purcell factor

$$F_p = 3Q\lambda^3/4\pi^2 n^3 V \quad [3]$$

Q is the quality factor of the combined quantum dot-cavity system, given by Q = $\lambda_{peak}/(\Delta\lambda_{cav} + \Delta\lambda_{QD})$, where $\Delta\lambda_{cav}$ and $\Delta\lambda_{QD}$ are the homogeneous line widths of the cavity mode and the quantum dot emission respectively[27,28]. In these room temperature experiments, $\Delta\lambda_{cav} \approx 0.1$ nm and $\Delta\lambda_{QD} \approx 14$ nm, so the relevant Q factor is approximately 45, determined by the exciton dephasing rate of the quantum dots. The Purcell factor for each confined mode therefore scales as $1/V$. If one assumes that the free space emission is unperturbed by the presence of the cavity then the modified decay rate is equal to $F_P + 1$. The mode volume dependence of this modified rate is shown as a solid line in figure 3(b).

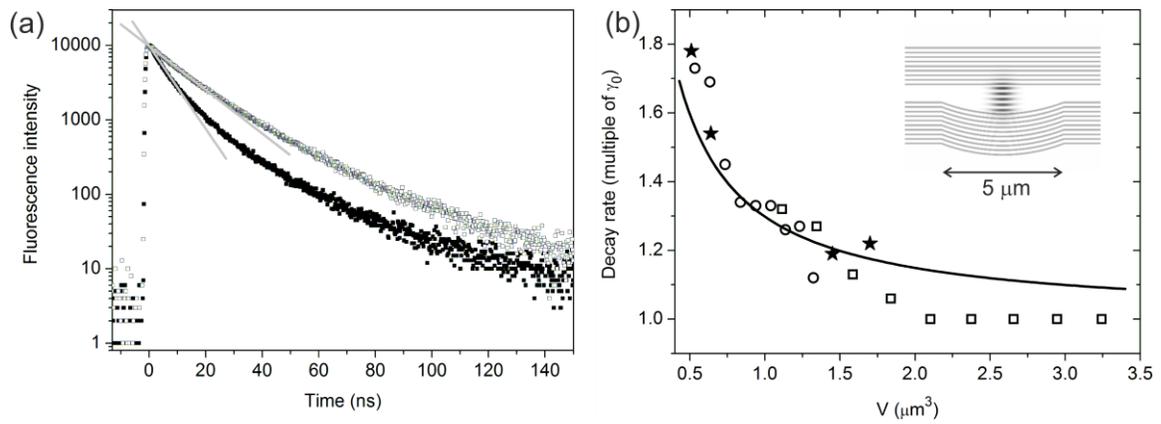

Fig. 3 (a) Semi-logarithmic plots of time resolved PL data measured from nanocrystal quantum dot ensembles outside the cavity (open shapes) and inside the smallest cavity (solid shapes). Lifetimes are obtained by fitting to the first 20 ns of the decay, indicated by solid grey lines. (b) Purcell enhancement as a function of mode volume for NQDs in cavities. Open shapes are derived from experimental fluorescence decay curves and the solid line is calculated using the Purcell equation 3. Circles correspond to $\beta = 7$ μm while squares correspond to $\beta = 25$ μm. Mode volumes are calculated from equation 2 in the text. The solid stars are wholly established from FDTD calculations. The inset shows a greyscale image of the cross sectional field distribution of the smallest cavity, as modeled using the FDTD method.

The simple model presented above can be developed to take account of three further physical factors: The influence of more than one resonant cavity mode; suppression of emission into unconfined continuum modes by the small cavities; and non-radiative recombination channels in the NQDs. Taking these factors into account leads to the expression

$$\gamma' = \gamma_{nr} + \left(\sum_i F_{P,i} + \alpha\right)\gamma_{rad} = \gamma_0\left(1 + \eta\left(\sum_i F_{P,i} + \alpha - 1\right)\right) \qquad [4]$$

where $\gamma_{nr}$ and $\gamma_{rad}$ are the non-radiative and radiative recombination rates of the NQDs in free space, $\gamma_0 = \gamma_{rad} + \gamma_{nr}$ is the total free space recombination rate, and $\eta = \gamma_{rad}/\gamma_0$ is the fluorescence quantum yield in free space. The variable α is the effective Purcell factor for emission into continuum modes ($0 \leq \alpha \leq 1$), representing the suppression of emission that is not matched to the confined modes of the cavity.

The modified spontaneous emission rates were also simulated numerically using FDTD modeling, by comparing the total power radiated by the source in the cavity to that when emitting into free space.[29,30] This method has the advantage of taking into account in detail the modified local density of states, thus calculating the contribution from both resonant cavity modes and leaky modes, corresponding to $\sum_i F_{P,i} + \alpha$ in equation (4). The source is constructed with its peak wavelength resonant with the strongest cavity mode and its line width equal to the homogeneous line width of the nanocrystals. This 'total power radiated' method yields, with no variable parameters, the four points marked as black stars in figure 3(b). These simulated rates agree remarkably well with the experimental data, with the modelled rate for the smallest mode volume equal to 1.78 $\gamma_0$.

The FDTD calculations allow us to isolate the contributions to the total emission rate as described in equation (4). For the smallest cavity we find that emission into the primary mode occurs at a rate close to that of the entire free space emission, with $F_P = 0.77$, and that emission into the continuum modes is suppressed by 32% giving α = 0.68. A small additional coupling to other cavity modes within the NQD line width makes up the remainder of the total emission rate. We now turn our attention to the cavity coupling efficiency of the emission, for which the quantum efficiency of the NQD also comes into consideration. The close agreement between the measured lifetimes and those modeled using FDTD support the claim that $\eta$ is close to unity, but we continue to include the parameter here for completeness. It can easily be seen from equation 4 that the quantum efficiency for coupling into the $i^{th}$ cavity mode is given by

$$\eta_i = \frac{\eta F_{P,i}}{1+\eta(\sum_i F_{P,i}+\alpha-1)} \qquad [5]$$

Using the values determined above and $\eta = 0.9$ we find that emission into the primary cavity mode occurs with an efficiency of 40.7%.

Finally, we demonstrate the measurement of emission from a single NQD into the cavity mode. This is achieved simply by diluting the quantum dot solution sufficiently that on average a single dot is present in the cavity mode at any given time. The evidence for single dot emission is twofold. Firstly, Figure 4(a) shows the spectrum for a single NQD at room temperature in free space and compares this to the emission seen in the cavity. Despite the presence of cavity modes at about 8 nm intervals, the range of excited modes is comparable with the homogeneous line width of 14 nm, in contrast with the spectra in Figure. 2(b) and 2(d) in which emission is observed across the inhomogeneously broadened ensemble spectrum. Secondly Figure 4(b) shows a time trace of the spectrum, which reveals substantial fluctuation in the mode intensity that is attributable to single dot fluorescence intermittency.

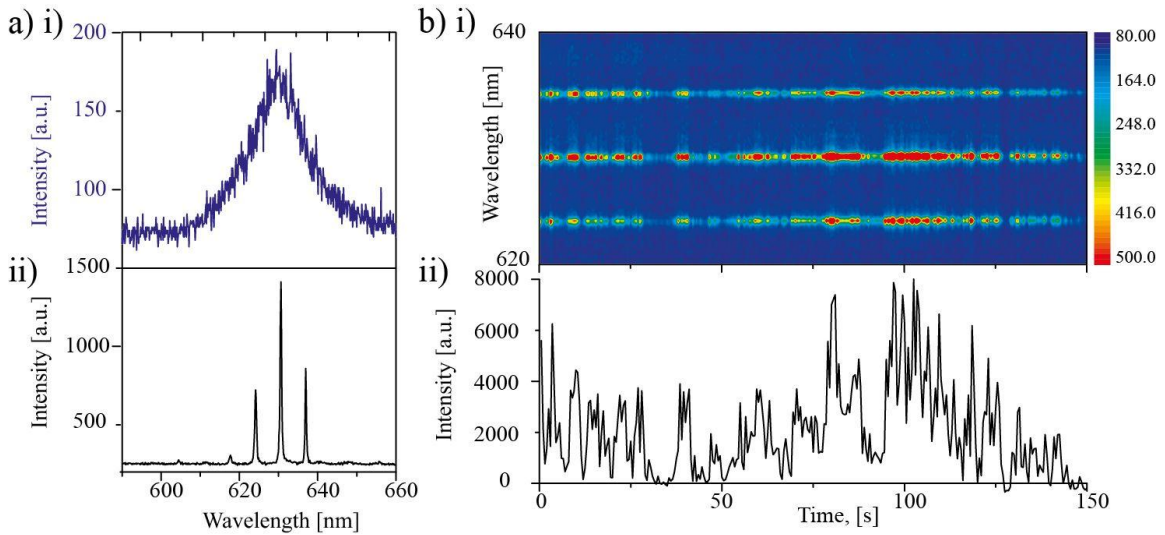

Figure 4: Emission from a single CdSe/ZnS NQD coupled to a cavity at room temperature. (a) Emission spectrum from a single NQD (i) with no cavity present, and (ii) in a cavity ($\beta = 7\mu m$, $L = 4\mu m$); (b) time trace of (i) the emission spectrum and (ii) the total emission intensity showing

characteristic blinking behavior indicative of a single emitter. The excitation intensity is 10 kW $cm^{-2}$ and the integration time of the detector is 0.5 s.

It is encouraging to note that the photon count rate from the single nanocrystal in the cavity is at least comparable to that measured in the absence of a cavity with an NA = 1.25 oil immersion lens. The latter would be expected to collect 35% of the light emitted on average from a single NQD, providing broad agreement with the modelled quantum efficiency calculated using FDTD.

In summary, the spontaneous emission of CdSe/ZnS quantum dots in arrays of tunable open-access optical microcavities at room temperature is reported. We demonstrate efficient coupling of spontaneous emission into resonant cavity modes, and output coupling into low numerical aperture external optics. A spontaneous emission rate enhancement is observed due to the Purcell effect and effectively modelled using FDTD simulations. We further demonstrate the measurement of emission from a single quantum dot coupled to cavity modes. Our results show significant promise for room temperature single photon sources, and for using these microcavities in single molecule fluorescence sensing applications.

**Acknowledgments**

The authors acknowledge funding from EPSRC. Z-YD acknowledges support from the K C Wong foundation.  JMS acknowledges additional support from Hewlett Packard Ltd. We would also like to thank eBioscience Inc. for supplying the NQD samples.